\documentclass[aps,prl,reprint,superscriptaddress]{revtex4-2}
\usepackage{amsmath,amssymb,amsthm,mathtools,bm}
\usepackage{bbm}
\usepackage{physics}
\usepackage{microtype}

\newcommand{\cH}{\mathcal H}
\newcommand{\cN}{\mathcal N}
\newcommand{\id}{\mathbbm{1}}

\begin{document}

\title{Relativistic Covariance and Nonlinear Quantum Mechanics: Tomonaga-Schwinger Analysis}

\author{Stephen D.H. Hsu}
\affiliation{Department of Physics and Astronomy, Michigan State University}

\begin{abstract}
We use the Tomonaga--Schwinger (TS) formulation of quantum field theory to determine when state-dependent additions to the local Hamiltonian density (i.e., modifications to linear Schr\"odinger evolution) violate relativistic covariance. We derive new operator integrability conditions required for foliation independence, including the Fr\'echet derivative terms that arise from state-dependence. Nonlinear modifications of quantum mechanics affect operator relations at spacelike separation, leading to violation of the integrability conditions.
\end{abstract}

\maketitle

\section{Tomonaga--Schwinger formulation}

The linear structure of quantum mechanics has deep consequences, such
as the persistence of superpositions. Is this linearity fundamental, or merely an approximation? 

The central question addressed in this work is whether nonlinear 
(state-dependent) modifications to quantum field theory can be made 
compatible with relativistic covariance. Models in 
non-relativistic quantum mechanics often include, for example, “instantaneous” potentials such as the Coulomb potential. However, quantum field theories used in particle physics are known to describe local physics with relativistic causality (influences propagate only within the light cone), making violations of this property easier to identify.

Here we examine the requirement of foliation independence -- that physical 
predictions should not depend on the choice of spacelike slicing. In the Tomonaga-Schwinger formalism, which provides a covariant  description of quantum evolution, this requirement leads to a precise  integrability condition.

Let $\Sigma$ be a spacelike Cauchy surface, and $\ket{\Psi,\Sigma}$ the state on it.
In linear local QFT the Tomonaga--Schwinger (TS) equation is
\begin{equation}
  i\hbar\,\frac{\delta}{\delta\sigma(x)}\ket{\Psi,\Sigma} = \hat{\cH}(x)\ket{\Psi,\Sigma},
  \label{eq:TS-linear}
\end{equation}
where $\sigma(x)$ denotes deformation of the hypersurface in the normal direction at $x$. Foliation independence,
$\big[\delta/\delta\sigma(x),\delta/\delta\sigma(y)\big]\ket{\Psi, \Sigma}=0$ for spacelike $x,y$,
requires $[ \hat{\cH}(x), \hat{\cH}(y) ]=0$, the usual microcausality condition~\cite{Schwinger1951,DeWitt1967}.

\section{State-dependent modification}

We now allow a deterministic nonlinear term,
\begin{equation}
 i\hbar\,\frac{\delta}{\delta\sigma(x)}\ket{\Psi,\Sigma}
   = (\hat{\cH}(x)+\hat{\cN}_x[\Psi])\ket{\Psi,\Sigma},
   \label{eq:TS-nonlinear}
\end{equation}
where $\hat{\cN}_x[\Psi]$ is an operator-valued functional of the global state.
We assume differentiability of $\hat{\cN}_x[\Psi]$ on the projective Hilbert space and deterministic, norm-preserving evolution, as in Weinberg~\cite{Weinberg1989}. See also \cite{KaplanRajendran2022,BerglundGeraciHubschMattinglyMinic2023,ChodosCooper2025Generalized,RembielinskiCaban2020Nonlinear} for more recent discussion of nonlinear QM. Some proposals may not be of the form (\ref{eq:TS-nonlinear}) and hence are beyond the scope of this paper.

For a map $\hat{\cN}_x:\mathsf{X}\to\mathcal{B}(\mathsf{X})$ (bounded functions on the state space), its Fr\'echet derivative at $\ket{\Psi}$ is
\begin{equation}
 D\hat{\cN}_x|_{\Psi}[\delta\Phi]
  = \lim_{\epsilon\to0}\frac{\hat{\cN}_x[\Psi+\epsilon\delta\Phi]-\hat{\cN}_x[\Psi]}{\epsilon},
\end{equation}
so a local deformation at $y$ induces
\begin{equation}
 \delta_y\hat{\cN}_x
 = D\hat{\cN}_x|_{\Psi}
    \!\left[\frac{\delta\ket{\Psi}}{\delta\sigma(y)}\right]
 = -\frac{i}{\hbar}\,
   D\hat{\cN}_x|_{\Psi}
    \!\left[(\hat{\cH}(y)+\hat{\cN}_y[\Psi])\ket{\Psi}\right].
 \nonumber
\end{equation}
In words, this is the change in the nonlinear operator $\hat{\cN}$ at $x$ due to the change in state $\Psi$ caused by a deformation at $y$.
\section{Integrability identity}

Applying the mixed functional derivatives and subtracting gives
\begin{align}
  0 &=
  \Big[\frac{\delta}{\delta\sigma(x)},\frac{\delta}{\delta\sigma(y)}\Big]\ket{\Psi} \\
  &= \frac{-1}{\hbar^2}\!
     \Big(
        [\hat{\cH}(x){+}\hat{\cN}_x,\hat{\cH}(y){+}\hat{\cN}_y]
        + i\hbar(\delta_y \hat{\cN}_x - \delta_x \hat{\cN}_y)
     \Big)\ket{\Psi} \nonumber
\end{align}
which yields the state-dependent operator constraint
\begin{align} 
  [\hat{\cH}(x),\hat{\cN}_y]
  + [\hat{\cN}_x,\hat{\cH}(y)]
  &+ [\hat{\cN}_x,\hat{\cN}_y] \nonumber \\
  &+ i\hbar(\delta_y \hat{\cN}_x - \delta_x \hat{\cN}_y)=0 
  \label{eq:star}
\end{align} 
where $x,y$ are spacelike separated: $x\sim y$ (after smearing). Equation~\eqref{eq:star} must hold for all admissible states if the theory is to yield consistent, foliation-independent evolution.  It generalizes the standard TS integrability condition of Schwinger~\cite{Schwinger1951} and DeWitt~\cite{DeWitt1967}.

Note that (\ref{eq:star}) is expressed as an operator equation, with operators that are functions of spacetime coordinates $x$ and $y$. However, the situation under state-dependent evolution is much more complex than in the usual linear quantum mechanics, because $\hat{\cN}_x$ is explicitly state-dependent, and any time-evolution operator used to relate operators at different times is also state-dependent. We will return to this point below. 

\section{Two-bubble composition check}

To verify \eqref{eq:star} directly, consider infinitesimal deformations at $x$ and $y$:
\begin{equation}
  U_x(\epsilon) = \id - \frac{i\epsilon}{\hbar}(\hat{\cH}(x)+\hat{\cN}_x[\Psi])
  + \mathcal{O}(\epsilon^2).
\end{equation}
Sequentially applying $U_x$ and $U_y$ and expanding to $\mathcal{O}(\epsilon^2)$ (formally, on $\mathcal{D}$) yields
\begin{align}
  (U_x & U_y - U_yU_x) \ket{\Psi} = \nonumber \\
  & \frac{\epsilon^2}{\hbar^2}\!\left(
      [\hat{\cH}(x){+}\hat{\cN}_x,\hat{\cH}(y){+}\hat{\cN}_y]
      + i\hbar(\delta_y\hat{\cN}_x-\delta_x\hat{\cN}_y)
    \right)\!\ket{\Psi} ~, \nonumber
\end{align}
confirming that this state-dependent operator equation is a necessary condition for hypersurface-independent TS evolution on the set of admissible states.

\section{Example: Weinberg operator-expectation nonlinearity}
\label{sec:weinberg}

We now examine the Weinberg-type nonlinearity \cite{Weinberg1989}
\begin{equation}
  \hat{\cN}_x[\Psi] = \lambda\,\langle \Psi | \hat O(x) | \Psi \rangle\,\hat O(x),
  \label{eq:weinberg}
\end{equation}
where $\hat O(x)$ is a local Hermitian operator density and $\lambda\in\mathbb{R}$.  
The state-dependence in this case is rather mild: it enters via the expectation value of the operator $\hat O(x)$. This allows explicit TS computations resulting in seemingly well-defined quantities modulo the question of how operators themselves evolve in time under state-dependent dynamics. We return to the latter question in the next section.

Norm preservation: If $\hat O(x)$ is Hermitian and $\lambda$ real, then $\hat{\cN}_x[\Psi]$ is Hermitian for each fixed $\Psi$, so the total Tomonaga–Schwinger generator $\hat{\cH}(x)+\hat{\cN}_x[\Psi]$ remains symmetric, and norm is preserved automatically.

\paragraph{Fr\'echet derivative.}
For a small displacement $\ket{\Psi}\mapsto \ket{\Psi}+\ket{\delta\Phi}$,
\begin{equation}
  \delta\langle\hat O(x)\rangle_\Psi
  = \bra{\delta\Phi}\hat O(x)\ket{\Psi}
  + \bra{\Psi}\hat O(x)\ket{\delta\Phi}.
\end{equation}
Hence the Fr\'echet derivative acts by
\begin{equation}
  D\hat{\cN}_x|_{\Psi}[\delta\Phi]
  = \lambda\Big(
    \bra{\delta\Phi}\hat O(x)\ket{\Psi}
    + \bra{\Psi}\hat O(x)\ket{\delta\Phi}
  \Big)\hat O(x) \nonumber
  \label{eq:DN-general}
\end{equation}
which equals $2\lambda\,\mathrm{Re}\,\bra{\delta\Phi}\hat O(x)\ket{\Psi}$ times $\hat O(x)$ and is therefore Hermitian.

Finite-dimensional example: Consider a single qubit with $\hat O=\sigma_z=\mathrm{diag}(1,-1)$ and an unnormalized state $\ket{\Psi}=(a,b)^{\top}$. Then $\langle \hat O\rangle_\Psi=\lvert a\rvert^2-\lvert b\rvert^2$. Under $\ket{\Psi}\mapsto \ket{\Psi}+\epsilon\ket{\delta\Phi}$ with $\ket{\delta\Phi}=(\delta a,\delta b)^{\top}$, the first-order change is
$
\delta\langle \hat O\rangle
=\bra{\delta\Phi}\sigma_z\ket{\Psi}+\bra{\Psi}\sigma_z\ket{\delta\Phi}.
$

\paragraph{Directional variation along a local TS deformation.}
Using the nonlinear TS equation,
\begin{equation}
  \frac{\delta\ket{\Psi}}{\delta\sigma(y)}
  = -\frac{i}{\hbar}\big(\hat{\cH}(y)+\hat{\cN}_y[\Psi]\big)\ket{\Psi}
  = -\frac{i}{\hbar}\hat G(y)\ket{\Psi},
\end{equation}
with $\hat G(y):=\hat{\cH}(y)+\hat{\cN}_y[\Psi]$ Hermitian, we find
\begin{align}
\delta_y\hat{\cN}_x
 &= D\hat{\cN}_x|_{\Psi}\!\left[\frac{\delta\ket{\Psi}}{\delta\sigma(y)}\right] \nonumber \\
  &= \lambda\Bigg(
     \frac{i}{\hbar}\bra{\Psi}\hat G(y)\hat O(x)\ket{\Psi}
     -\frac{i}{\hbar}\bra{\Psi}\hat O(x)\hat G(y)\ket{\Psi}
   \Bigg)\hat O(x) \nonumber\\[4pt]
 &= \frac{i\lambda}{\hbar}\,\langle[\hat G(y),\hat O(x)]\rangle_\Psi\,\hat O(x).
 \label{eq:weinberg-delta-correct}
\end{align}

\paragraph{Evaluation of the integrability condition.}
Substituting \eqref{eq:weinberg} and \eqref{eq:weinberg-delta-correct} into the TS integrability identity~\eqref{eq:star}:
\begin{align}
[\hat{\cH}(x),\hat{\cN}_y]
 &= \lambda\,\langle \hat O(y)\rangle_\Psi\, [\hat{\cH}(x),\hat O(y)],\nonumber\\
[\hat{\cN}_x,\hat{\cH}(y)]
 &= \lambda\,\langle \hat O(x)\rangle_\Psi\, [\hat O(x),\hat{\cH}(y)],\nonumber\\
[\hat{\cN}_x,\hat{\cN}_y]
 &= \lambda^2\,\langle \hat O(x)\rangle_\Psi\langle \hat O(y)\rangle_\Psi\, [\hat O(x),\hat O(y)].
\end{align}
The the integrability conditions are related to operator commutation relations for spacelike $x\sim y$. Under the usual assumption of microcausality  (which we will revisit below!), all three commutators vanish after smearing, and
\[
[\hat G(y),\hat O(x)]
  = [\hat{\cH}(y),\hat O(x)] + \lambda\langle\hat O(y)\rangle_\Psi[\hat O(y),\hat O(x)]
  = 0.
\]
Thus $\delta_y\hat{\cN}_x = 0$,  $\delta_x\hat{\cN}_y = 0$
and the cross-term
$C_4 = i\hbar(\delta_y\hat{\cN}_x-\delta_x\hat{\cN}_y)=0$.
Hence all contributions $C_1=C_2=C_3=C_4=0$, and the Weinberg form is consistent with relativistic foliation independence, {\it conditional on the assumption that operators commute at spacelike separation (microcausality)}.
However, as we discuss next, the assumption of microcausality cannot be consistently maintained under state-dependent evolution.

\section{Breakdown of Microcausality under Nonlinear Evolution}
\label{sec:microcausality-breakdown}

The condition of microcausality,
\begin{equation}
  [\hat O(x),\hat O'(y)] = 0 \qquad (x\sim y),
  \label{eq:microcausality}
\end{equation}
where $\hat O, \hat O'$ represent any operators, including the Hamiltonian density or Weinberg operator from the previous section, is usually imposed as an operator identity, reflecting the principle that local observables associated with spacelike-separated regions should commute.  In standard (linear) quantum field theory, this relation is preserved under unitary time evolution because the dynamics are implemented by a state-independent unitary map
\begin{equation}
  \hat O(x,t) = U^\dagger(t,t_0)\,\hat O(x,t_0)\,U(t,t_0) \nonumber
\end{equation}
\begin{equation}
  U(t,t_0) = \exp\!\left[-\frac{i}{\hbar}\!\int_{t_0}^t\! d^3z\,\hat{\cH}(z)\right]
  \label{eq:heisenberg}
\end{equation}
and $U(t,t_0)$ acts as an automorphism on the local operator algebra.  
Consequently, if~\eqref{eq:microcausality} holds on one spacelike surface, it continues to hold everywhere: linear unitarity transports the commutation relations consistently throughout spacetime.

For example, in canonical quantization, microcausality arises as a consequence of the
equal-time commutation relations imposed on the fundamental field operators
and their conjugate momenta.  
For a scalar field $\phi(\mathbf{x},t)$ with conjugate momentum
$\pi(\mathbf{x},t)$, the canonical commutation relations at an initial time
(e.g.\ $t=0$) are
\begin{align}
  [\,\phi(\mathbf{x},t),\,\phi(\mathbf{x}',t)\,] &= 0, \nonumber\\
  [\,\pi(\mathbf{x},t),\,\pi(\mathbf{x}',t)\,] &= 0, \nonumber \\
  [\,\phi(\mathbf{x},t),\,\pi(\mathbf{x}',t)\,] &= i\hbar\,\delta^{(3)}(\mathbf{x}-\mathbf{x}'). 
  \label{eq:CCR3}
\end{align}
Time evolution is generated by the Hamiltonian $\hat H = \int d^3x\,\hat{\cH}(x)$,
so that the equal-time commutation relations are preserved in time. Hence \eqref{eq:CCR3} remain valid for all $t$. A Lorentz boost can transform two spacelike-separated points
$x$ and $y$ into a reference frame in which they share the same time coordinate.  Consequently, for any pair of spacelike-separated spacetime points $x$ and $y$,
the field operators commute: $[\,\phi(x),\,\phi(y)\,] = 0 .$
This expresses the condition of
\emph{microcausality} in its canonical form:
observables associated with spacelike-separated regions cannot influence one
another, ensuring the consistency of relativistic locality in the operator
formalism.

\paragraph{State-dependent evolution.}
In a nonlinear or state-dependent theory \eqref{eq:TS-nonlinear} replaces the linear Schr\"odinger equation.  
The corresponding evolution from one hypersurface $\Sigma_0$ to another $\Sigma$ can no longer be represented by a fixed unitary operator $U(\Sigma,\Sigma_0)$ acting uniformly on the Hilbert space.  
Instead, the evolution depends on the current state $\ket{\Psi}$ itself, and one must write schematically
\begin{equation}
  \ket{\Psi,\Sigma} = U[\Psi;\Sigma,\Sigma_0]\,\ket{\Psi,\Sigma_0},
\end{equation}
where $U[\Psi;\Sigma,\Sigma_0]$ is a nonlinear, state-dependent transformation.
Because this transformation is not unitary in the usual sense, it does \emph{not} act as an algebra automorphism.
Consequently, the Heisenberg relation~\eqref{eq:heisenberg} no longer defines a consistent operator evolution, and there is no guarantee that commutation relations among spacelike-separated operators are preserved. Thus even if the microcausality condition~\eqref{eq:microcausality} is enforced on an initial hypersurface, it need not hold at later times.

\paragraph{Relation to the Ho--Hsu and Gisin--Polchinski analyses.}
Ho and Hsu~\cite{HoHsu2014} explicitly demonstrated that nonlinear
(state-dependent) modifications to the functional Schr\"odinger equation for quantum fields cause instantaneous entanglement between initially
unentangled subsystems $A$ and $B$ at spacelike separation. Specifically, \cite{HoHsu2014} considered two coherent wave packets localized in
spacelike-separated regions $A$ and $B$, initially in an unentangled product state
\begin{equation}
  \ket{\Psi(0)} = \ket{\psi_A}\!\otimes\!\ket{\psi_B}.
\end{equation}
Under a deterministic, state-dependent nonlinear evolution of the form
\begin{equation}
  i\hbar\,\partial_t \ket{\Psi(t)} = 
  \big(\hat H + \hat{\cN}[\Psi(t)]\big)\ket{\Psi(t)},
\end{equation}
they found that at the next instant of time, $t=0^+$, the global state becomes
entangled even though the regions $A$ and $B$ remain spacelike separated.
This instantaneous generation of correlations signals a breakdown of
operational locality, or equivalently, a failure of foliation independence in
the Tomonaga--Schwinger formulation.

For the Weinberg-type nonlinearity, given an initially factorized state $\Psi=\psi_A\otimes\psi_B$, separability at equal time is preserved only if
$[\hat \cH_A,\hat O_A]=0$ and $[\hat \cH_B,\hat O_B]=0$. If these commutators are nonzero, local expectation values such as
$\langle \hat O_B \rangle_{\psi_B}$ evolve in time,
and because each subsystem’s nonlinear term depends on the global (hence the other’s) expectation value,
the two regions become instantaneously entangled even at spacelike separation. If $[ \hat{\cH}, \hat{O} ] = 0$ the Weinberg modification to Schrödinger evolution becomes simply a shift in the Hamlitonian ($\langle \hat{O} \rangle$ is constant), and dynamics reverts to that of linear quantum mechanics.

In relation to microcausality, these results imply that for even for spacelike separated $A,B$: 
$\hat{O}_A \hat{O}_B \vert \psi \rangle \neq \hat{O}_B \hat{O}_A \vert \psi \rangle .$
The operator acting at $B$ instantaneously changes the system at $A$ (via state dependence), and vice-versa, so that 
$$ \hat{O}_A ( \hat{O}_B \vert \psi \rangle ) = \hat{O}_A \vert \psi'_B \rangle
\neq \hat{O}_B \vert \psi'_A \rangle = \hat{O}_B ( \hat{O}_A \vert \psi \rangle )~.$$

Gisin~\cite{Gisin1990} and Polchinski~\cite{Polchinski1991} reached the same conclusion in a nonrelativistic setting: deterministic nonlinear evolution of entangled pure states leads to superluminal signaling because the nonlinear dynamics couple spacelike-separated subsystems through the global wavefunction.

\paragraph{Implications for TS integrability.}

In a linear local theory, microcausality ensures foliation independence.  
However, when nonlinear (state-dependent) modifications of  are allowed, the TS integrability conditions are modified, and furthermore the
time evolution operator is not an algebra automorphism, so commutation relations at spacelike separation are not preserved. The integrability condition cannot hold for all states unless microcausality is dynamically enforced, which it is not.


\section{Weinberg nonlocal mean-field nonlinearity}
\label{sec:weinberg-nonlocal}

We now generalize Weinberg’s ansatz to a \emph{nonlocal mean-field} coupling:
\begin{equation}
  \hat{\cN}_x[\Psi]
  = \lambda\!\int\!d^3y\; f(x,y)\,
     \langle \Psi | \hat O(y) | \Psi \rangle\,\hat O(x),
  \label{eq:weinberg-nonlocal}
\end{equation}
where $f(x,y)$ is a smooth kernel of spacelike width (for example, a Gaussian falloff).  
In this model the generator at each point $x$ depends on the expectation values of field operators at \emph{all} points $y$, producing an explicitly nonlocal nonlinear term from the outset.

\paragraph{Violation of TS integrability.}
Because $\hat{\cN}_x[\Psi]$ depends on expectation values at all points $y'$, the Fr\'echet derivative acquires nonlocal contributions. The variation $\delta_y\hat{\cN}_x$ (from a deformation at $y$) is
\begin{equation}
\delta_y\hat{\cN}_x
 = \frac{i\lambda}{\hbar}\!\int\!d^3y'\,
 f(x,y')\,\big\langle[\hat G(y),\hat O(y')]\big\rangle_\Psi\,\hat O(x).
\end{equation}
Even assuming microcausality, the commutator $[\hat G(y),\hat O(y')]$ is supported where $y'$ lies in the (smeared) neighborhood of $y$. Thus the integral localizes (schematically, $\big\langle[\hat G(y),\hat O(y')]\big\rangle_\Psi \propto \delta^3(\vec y-\vec y')\,\langle \hat C(y)\rangle_\Psi$ for some local $\hat C(y)$), and is generically nonzero. The Fr\'echet cross-term in the integrability condition~\eqref{eq:star} is, for spacelike $x\sim y$, a generic state $\Psi$, and nontrivial kernel $f$, nonzero. Thus $C_4\neq 0$, and the integrability condition~\eqref{eq:star} fails.

\paragraph{Physical interpretation.}
This example illustrates a distinct mechanism for failure of the TS conditions: here, the violation occurs \emph{irrespective of microcausality} of the underlying operator algebra.  
Even if one imposes $[\hat O(x),\hat O(y)]=0$ for spacelike $x,y$, the nonlocal dependence of $\hat{\cN}_x[\Psi]$ on $\langle \hat O(y)\rangle_\Psi$ ensures that a deformation of the hypersurface near $y$ immediately alters the generator at distant $x$.  
The problem arises not from any breakdown of local commutation relations, but from the explicit functional coupling across spacelike regions encoded in $f(x,y)$.

\paragraph{Relation to wavefunction collapse.}
Such explicit nonlocality is necessary in deterministic nonlinear models that aim to describe wavefunction collapse or the suppression of macroscopic superpositions. Any nonlinear term capable of correlating distant degrees of freedom strongly enough to suppress macroscopic superpositions must itself depend on large-scale features of the state, thereby introducing nonlocal evolution.

\section{Retarded Nonlinearity in the Kaplan–Rajendran Model}

Kaplan and Rajendran (KR)~\cite{KaplanRajendran2022} proposed a nonlinear extension of quantum mechanics in which the evolution of a quantum field depends on the \emph{retarded} expectation values of operators in the past light cone of each spacetime point. The nonlinear correction to the Hamiltonian density is also of the mean field type studied above, but with a retarded kernel.
\begin{equation}
\hat N_x[\Psi]
= \int d^4x_1 \;
G_R(x;x_1)\,
\langle \hat O(x_1)\rangle_{\Psi} \, \hat P(x)
\end{equation}
where $G_R(x;x_1)$ is a retarded Green’s function and $\hat O, \hat P$ are field operators. The authors emphasize that this formulation guarantees \emph{causal propagation}—that is, no signal or influence travels outside the light cone—while maintaining unitarity and norm preservation.

However, this nonlinear modification does not satisfy the TS conditions. We obtain
\begin{align}
\delta_y\hat N_x
&=\frac{i}{\hbar}
\int d^4x_1\,G_R(x;x_1)
\left\langle
\big[\hat G(y), \hat O (x_1) \big]
\right\rangle_{\Psi}
\hat{P}(x),
\nonumber
\end{align}
where $\hat G(y)\;=\;\hat H(y)+\hat N_y[\Psi]$. The integrand has support where both factors are nonvanishing:
\begin{equation}
x_1 \in \text{supp}\, G_R(x;\cdot) \;\cap\;
\text{supp}\,\big\langle[\hat G(y),\hat O(x_1)]\big\rangle_{\Psi}.
\label{eq:support_intersection}
\end{equation}
Since $G_R(x;x_1)$ has support only for $x_1\in J^-(x)$ (the past light cone of $x$), 
and the commutator $\big[\hat G(y),\hat O(x_1)\big]$ is supported for $x_1\in J^-(y)$,
the intersection is
\begin{equation}
x_1 \in J^-(x)\cap J^-(y).
\label{eq:intersection_region}
\end{equation}
For two spacelike-separated points $x$ and $y$ on the same hypersurface, this intersection is generically nonempty.  
Thus, even though both kernels are individually causal, their combined dependence on the overlapping causal past of $x$ and $y$ yields a nonvanishing result for $\delta_y\hat N_x$, $\delta_x \hat N_y$, and their difference (i.e., $C_4$).

The TS integrability condition fails for generic states even though the KR model enforces retarded (causal) dependence.

\section*{Conclusion}

We have derived necessary conditions for relativistic covariance in state-dependent nonlinear extensions of quantum field theory using the Tomonaga-Schwinger framework. The key requirement is the integrability condition given in equation (\ref{eq:star}), which must hold for all admissible states to ensure foliation independence. This condition extends the standard microcausality requirement through additional terms arising from the Fréchet derivatives of the state-dependent contributions of the type considered in equation (\ref{eq:TS-nonlinear}).

While certain nonlinearities like the local Weinberg form may formally satisfy these conditions when microcausality is assumed, we have shown that state-dependent evolution fundamentally alters the behavior of operators at spacelike separation. The nonlinear dynamics generically leads to instantaneous entanglement generation between spacelike-separated regions, violating operational locality. This breakdown occurs because state-dependent evolution does not preserve operator commutation relations, even if microcausality is imposed on an initial Cauchy slice.

\section{Acknowledgements}

The author used AI models GPT-5, Gemini 2.5 Pro, and Qwen-Max in the preparation of this manuscript, primarily to check results, format latex, and explore related work. The author has checked all aspects of the paper and assumes full responsibility for its content.

\end{document}